\documentclass{article}
\usepackage{makeidx}
\usepackage{amsfonts}
\usepackage{amsmath}
\usepackage{epsfig}
\usepackage{latexsym}
\usepackage{amssymb}
\usepackage{graphicx}

\newtheorem{theorem}{Theorem}

\newtheorem{lemma}[theorem]{Lemma}
\newtheorem{proposition}[theorem]{Proposition}

\begin{document}

\title{On singular perturbations of quantum dynamical semigroups}
\author{A. S. Holevo\\
Steklov Mathematical Institute \\
of Russian Academy of Sciences, Moscow}
\date{}
\maketitle

\begin{abstract}
We consider two examples of dynamical semigroups obtained by
singular perturbations of a standard generator which are special case of
unbounded completely positive perturbations studied in detail in \cite{hol4}. In the section \ref{s1} we propose a generalization of an example from \cite{arv} aimed to give a positive answer to a conjecture of Arveson. In the section \ref{s2} we consider in
greater detail an improved and simplified construction of a nonstandard dynamical semigroup outlined
in our short communication \cite{hol5}.
\end{abstract}

\section{Introduction. Standard dynamical semigroups}

 Let $\mathcal{H}$ be a separable Hilbert space, $\mathfrak{L}(%
\mathcal{H})$ the algebra of all bounded operators and $\mathfrak{T}(%
\mathcal{H})=\mathfrak{L}(\mathcal{H})_{\ast }$ the Banach space of
trace-class operators in $\mathcal{H}.$ Quantum dynamical semigroup in $%
\mathfrak{L}(\mathcal{H})$ is a one-parameter semigroup $%
T_{t},\,t\geq 0,$ of normal completely positive (CP) maps of $\mathfrak{L}(%
\mathcal{H})$ such that $T_{0}=\mathrm{Id}$ (the identity map),
and $T_{t}[X]$ is continuous in $t$ for any fixed $X\in \mathfrak{L}(
\mathcal{H})$ in the weak operator topology. There is a unique preadjoint
semigroup $S_{t}=\left( T_{t}\right) _{\ast }$ in $\mathfrak{T}(\mathcal{H}). $
It was shown in the seminal paper \cite{lindblad} that $S_{t}$ is
norm continuous semigroup if and only if its generator is bounded and has
the representation
\begin{equation}
\mathcal{K}[\omega ]=\sum_{j}L_{j}\omega L_{j}^{\ast }-\omega K^{\ast
}-K\omega ,\quad \omega \in \mathfrak{T}(\mathcal{H}),  \label{0}
\end{equation}%
where the series $\sum_{j}L_{j}^{\ast }L_{j}$ strongly converges (in the
finite dimensional case the characterization of the generator was obtained
in \cite{gks}). In this case the conservativity condition $%
\sum_{j}L_{j}^{\ast }L_{j}=K+K^{\ast }$ implies that the semigroup $S_{t}$
is trace-preserving while $T_{t}$ is unital. In the case of non-norm
continuous semigroups, corresponding to unbounded generators, this is no
longer the case, and additional conditions are required to ensure the trace
preservation. This is a quantum analog of the explosure or extinction (absorption)
phenomena for classical Markov processes, see e.g. \cite{fel}, \cite{chebot1}. Also a
question of universality of a representation of the type (\ref{0}) arizes
which will be the main concern of the present note, see also \cite{hsw}.

Quantum dynamical semigroups with unbounded generators were considered in
\cite{dav2, chebot, chebot1, chf, sinha, bhat, hol3, hol4, hol5}, see also
references therein. We will refer in particular to the Appendix in \cite{hol9}
where the backward and forward quantum Markovian master equations (MME) with
unbounded generators are described. Let $K,L_{j}$ be linear operators
defined on a dense domain $\mathcal{D}$ of a Hilbert space $\mathcal{H}$ ,
satisfying the dissipativity condition%
\begin{equation}
\sum_{j}\left\Vert L_{j}\psi \right\Vert ^{2}\leq 2\mathrm{Re}\,\langle \psi
|K|\psi \rangle ,\quad \psi \in \mathcal{D},  \label{1}
\end{equation}%
in paricular, $K$ is accretive, $\mathrm{Re}\,\langle \psi |K|\psi \rangle
\geq 0$ . In case of equality in (\ref{1}) it is called the conservativity
condition. We assume \ that $K$ is maximal accretive ($m$-accretive)
operator and $\mathcal{D}$ is invariant subspace for the contraction
semigroup $\exp (-Kt),\,t\geq 0$, hence a core for $K$ (Theorem X.49 of \cite%
{rs2}) (when it is convenient, we can take$\mathcal{\ D}=\mathrm{dom}\,K$).
Then there exists the unique minimal solution $T_{t},\,t\geq 0,$ of the
Cauchy problem for the backward MME
\begin{eqnarray}
\frac{d}{dt}\langle \varphi |T_{t}[X]|\psi \rangle &=&\sum_{j}\langle
L_{j}\varphi |T_{t}[X]|L_{j}\psi \rangle -\langle K\varphi |T_{t}[X]|\psi
\rangle -\langle \varphi |T_{t}[X]|K\psi \rangle ,  \label{2} \\
\quad \varphi ,\psi &\in &\mathcal{D},\quad X\in \mathfrak{L}(\mathcal{H}),
\notag
\end{eqnarray}%
satisfying the condition $T_{0}[X]=X,$ which is a dynamical semigroup on the
algebra $\mathfrak{L}(\mathcal{H})$ of all bounded operators in $\mathcal{H}$
(see \cite{chebot1}, cf. also Theorem A.1 in \cite{hol9}).

Denoting by $\mathcal{K}$ the generator of the preadjoint semigroup $%
S_{t}=\left( T_{t}\right) _{\ast }$ in $\mathfrak{T}(\mathcal{H}),$ the MME (%
\ref{2}) can be rewritten in the form
\begin{equation}
\frac{d}{dt}\langle \varphi |T_{t}[X]|\psi \rangle =\mathrm{Tr\,}\mathcal{K\,%
}[|\psi \rangle \langle \varphi |\,]T_{t}[X],\quad \varphi ,\,\psi
\in \mathcal{D},  \label{2a}
\end{equation}%
where

\begin{equation}
\mathcal{K\,}[|\psi \rangle \langle \varphi |\,]=\sum_{j}|L_{j}\psi \rangle
\langle L_{j}\varphi |\,-\,|K\,\psi \rangle \langle \varphi |\,-|\,\psi
\rangle \langle K\varphi |\,.  \label{K1}
\end{equation}%
The dissipativity (\ref{1}) amounts to the inequality $\mathrm{Tr\,}\mathcal{K}[|\psi
\rangle \langle \psi |]\leq 0$ and conservativity -- to the equality $\mathrm{Tr\,}\mathcal{K%
}[|\psi \rangle \langle \psi |]=0,\,\psi \in \mathcal{D}$.

The dynamical semigroups $T_{t}$ and $S_{t}$ will be called \textit{standard}
if they can be obtained with the procedure described above.

In this paper we consider two examples of dynamical semigroups obtained by
singular perturbations of a standard generator which are special case of
unbounded completely positive perturbations studied in detail in \cite{hol4}.
In the section \ref{s1} we propose a generalization of an example in \cite%
{arv} aimed to give a positive answer to a conjecture of Arveson. It is
itself standard in the sense defined above although involves nonclosable
operators $L_{j}$ (cf. \cite{alaz}). In the section \ref{s2} we consider in
greater detail a construction of a nonstandard dynamical semigroup outlined
in the short communication \cite{hol5} following Example 3 in \cite{hol4}.
We provide an improvement and simplification of the argument in \cite{hol5},
including a minor correction.

The singular perturbations of the generator (\ref{K1}) we consider are rank
one perturbations of the form%
\begin{equation*}
\Lambda \lbrack \omega ]=-\Omega \,\mathrm{Tr\,}\mathcal{K}[\omega ],\quad
\omega \in \mathrm{dom\,}\mathcal{K\,},
\end{equation*}%
where $\Omega $ is a fixed density operator in $\mathcal{H}$ . In such case
the perturbed generator is conservative in the sense that%
\begin{equation*}
\mathrm{Tr\,}\left( \mathcal{K}[\omega ]+\Lambda \lbrack \omega ]\right)
=0\,,\quad \omega \in \mathrm{dom\,}\mathcal{K\,},
\end{equation*}%
moreover $\mathrm{dom}\left( \mathcal{K}+\Lambda \right) =\mathrm{dom\,}%
\mathcal{K\,},$ as shown in \cite{hol4}. It follows that the perturbed
dynamical semigroup $S_{t}$ is trace-preserving, resp. $T_{t}$ is unital.

\section{\ A generalization of Arveson's example}

\label{s1} In this example $\mathcal{H=}$ $L^{2}(\mathbb{R}_{+}),$ $K=-\frac{%
d}{dx},$ $\mathcal{D}=AC^{1}(\mathbb{R}_{+})$ is the subspace of absolutely
continuous functions $\varphi (x)$ on $\mathbb{R}_{+}$ such that $\varphi
^{\prime }(x)\in $ $L^{2}(\mathbb{R}_{+}).\ $ Thus$\,W_{t}=\exp (-Kt),t\geq
0,$ is the semigroup of left shifts in $\mathcal{H},$ $W_{t}\varphi
(x)=\varphi (x+t).$ Then $W_{t}^{\ast }W_{t}$ is a projector on the subspace
of functions vanishing on $[0,t]$ and $W_{t}W_{t}^{\ast }=I.$ 

In what follows we identify trace-class operators $
\omega $ in $\mathcal{H}$  with their kernels $\omega
(x,y),\,x,y\in \mathbb{R}_{+}.$  Consider the
quantum dynamical semigroup
\begin{equation}
S_{t}^{0}[\omega ]=W_{t}\omega W_{t}^{\ast };\quad S_{t}^{0}[\omega
](x,y)=\omega (x+t,y+t)  \label{S01}
\end{equation}%
acting on operators $\omega\in\mathfrak{T}(\mathcal{H}),$ with the generator
\begin{equation*}
\mathcal{K}_{0}\omega =  \omega'_x (x,y) + \omega'_y (x,y)=\frac{d}{dt}\omega (x+t,y+t)|_{t=0},
\end{equation*}%
defined initially on the domain
\begin{equation*}
\mathfrak{D}_{0}=\mathrm{lin}\left\{ \omega :\omega =|\varphi \rangle
\langle \psi |;\,\varphi ,\psi \in \mathcal{D}\right\} \,\mathfrak{\subset T}%
(\mathcal{H}).
\end{equation*}%
A closed description of $\mathcal{K}_{0}$ is given by the following
Proposition the proof of which is postponed until the end of this Section.

\begin{proposition}
\label{p1} The domain $\mathrm{dom\,}\mathcal{K}_{0}$ consists of trace-class operators
$\omega $ with kernels $\omega (x,y)$ such that for almost all $(x,y)$ the
function $\omega (x+t,y+t)$ as a function of $t$ is absolutely continuous and its derivative $\sigma (x,y)=\frac{d}{dt}\omega
(x+t,y+t)|_{t=0}$ is a kernel of a trace-class operator in $\mathcal{H}$.
\end{proposition}

Now consider the perturbed generator
\begin{equation}
\mathcal{K}[\omega ]=\mathcal{K}_{0}[\omega ]+\Lambda \lbrack \omega ],\quad
\quad \omega \in \mathrm{dom\,}\mathcal{K}_{0},  \label{K5}
\end{equation}%
where
\begin{equation*}
\Lambda \lbrack \omega ]=-\Omega \,\mathrm{Tr\,}\mathcal{K}_{0}[\omega
]=\Omega \,\omega (0,0),
\end{equation*}%
$\Omega $ is a density operator. Here the second equality follows from
\begin{equation*}
\int_{0}^{\infty }\frac{d}{dx}\omega (x,x)dx=-\,\omega (0,0).
\end{equation*}

Since the perturbation has rank one, $\mathrm{dom\,}\mathcal{K}=\mathrm{dom\,%
}\mathcal{K}_{0}$ \cite{hol4}. The perturbed (minimal) dynamical semigroup $%
S_{t}$ constructed as in \cite{hol4} is \textit{standard} and
trace preserving. Indeed, the dual semigroup $T_{t}=\left( S_{t}\right)
^{\ast }$ is unital and is the minimal solution of the backward MME (\ref{2a}),
which in our case reads
\begin{equation}
\frac{d}{dt}\langle \varphi |T_{t}[X]|\psi \rangle =\,\mathrm{Tr}\,\left(
\mathcal{K}_{0}+\Lambda \right) [|\psi \rangle \langle \varphi
|]T_{t}[X],\quad \varphi ,\psi \in \mathcal{D},  \label{bme}
\end{equation}%
where
\begin{equation*}
\mathcal{K}_{0}[|\psi \rangle \langle \varphi |]=|\psi ^{\prime }\rangle
\langle \varphi |+|\psi \rangle \langle \varphi ^{\prime }|=-|K\,\psi \rangle
\langle \varphi |\,-|\,\psi \rangle \langle K\varphi |;\end{equation*}
\begin{equation}\Lambda
\lbrack |\psi \rangle \langle \varphi |]=\Omega \,\psi (0)\,\overline{
\varphi (0)}=\sum_{j}|L_{j}\psi \rangle \langle L_{j}\varphi |.  \label{lambda}
\end{equation}
Here $L_{j}$ are \textit{nonclosable} operators (cf. \cite{alaz})
\begin{equation*}
L_{j}|\psi \rangle =|l_{j}\rangle \,\psi (0),\quad \psi \in \mathcal{D}%
=AC^{1}(\mathbb{R}_{+}),
\end{equation*}
and the vectors $|l_{j}\rangle$ are such that $\Omega =\sum_{j}|l_{j}\rangle \langle l_{j}|$.

In Arveson's example \cite{arv} $\Omega =|f\rangle \langle f|$, where $%
f(x)=c\exp (-\alpha x)$. In that case the perturbed semigroups are
constructed explicitly as%
\begin{eqnarray}
S_{t}[\omega ] &=&W_{t}\omega W_{t}^{\ast }+\mathrm{\,}\Omega \,\mathrm{Tr\,}%
\left( I-W_{t}^{\ast }W_{t}\right) \omega ,  \notag \\
T_{t}[X] &=&W_{t}^{\ast }XW_{t}+\left( I-W_{t}^{\ast }W_{t}\right) \,\mathrm{%
Tr\,}\Omega X.  \label{SA}
\end{eqnarray}%
Arveson introduces a general notion of the domain algebra of a quantum
dynamical semigroup with generator $\mathcal{L}$\ as
\begin{equation*}
\mathcal{A=}\left\{ X\in \mathrm{dom\,}\mathcal{L}:X^{\ast }X\in \mathrm{%
dom\,}\mathcal{L},\,\,XX^{\ast }\in \mathrm{dom\,}\mathcal{L}\right\}
\end{equation*}%
and shows that the strong closure of the domain algebra for the semigroup (%
\ref{SA}) consists of all operators commuting with $\Omega =|f\rangle
\langle f|.$ He also asks if there exist dynamical semigroups whose domain
algebra is as small as $\mathbb{C}\cdot I$ . Our generalization allows to
give a positive answer to this question.

\begin{theorem}
\label{t1} Let $\Omega $ be an arbitrary density operator, $T_{t}$ -- the
dynamical semigroup solving (\ref{bme}) and $\mathcal{L}$ -- its generator.
If $X\in \mathcal{A}$, the domain algebra of $T_{t},$ then
\begin{equation*}
\left[ X-\left( \mathrm{Tr\,}\Omega X\right) I\right] \Omega =\Omega \left[
X-\left( \mathrm{Tr\,}\Omega X\right) I\right] =0.
\end{equation*}%
In particular, if $\Omega $ is nondegenerate, then $X$ is a multiple of
identity, so that $\mathcal{A=}\mathbb{C}\cdot I$ .
\end{theorem}

We use the criterion (iii) of Lemma 1 in \cite{arv} for an arbitrary quantum
dynamical semigroup $T_{t}$ which says that $X\in \mathrm{dom\,}\mathcal{L}$
iff%
\begin{equation}
\sup_{t>0}t^{-1}\left\Vert T_{t}[X]-X\right\Vert \leq M<\infty .  \label{iii}
\end{equation}

\begin{lemma}
\label{l1} Let $\mathcal{L}_{0}$ be the generator of the semigroup $%
T_{t}^{0}=\left( S_{t}^{0}\right) ^{\ast },$ $T_{t}^{0}[X]=W_{t}^{\ast
}XW_{t}.$ Then $\mathrm{dom\,}\mathcal{L}_{0}$ is $\ast $-algebra and $%
I\notin \mathrm{dom\,}\mathcal{L}_{0}.$
\end{lemma}

\textit{Proof.} $W_{t}^{\ast }W_{t}$ is a projector, hence $\left\Vert
T_{t}^{0}[I]-I\right\Vert =\left\Vert W_{t}^{\ast }W_{t}-I\right\Vert =1,$
hence $I\notin \mathrm{dom\,}\mathcal{L}_{0}.$

The first assertion holds for any semigroup of endomorphisms, but we include
the proof for completeness. Apparently $\mathrm{dom\,}\mathcal{L}_{0}$ is
selfadjoint.

Using the identity $W_{t}W_{t}^{\ast }=I,$
\begin{equation*}
t^{-1}\left( T_{t}^{0}[XY]-XY\right) =t^{-1}\left( T_{t}^{0}[X]-X\right)
T_{t}^{0}[Y]+t^{-1}X\left( T_{t}^{0}[Y]-Y\right) ,
\end{equation*}%
hence%
\begin{equation*}
t^{-1}\left\Vert T_{t}^{0}[XY]-XY\right\Vert \leq t^{-1}\left\Vert
T_{t}^{0}[X]-X\right\Vert \left\Vert T_{t}^{0}[Y]\right\Vert
+t^{-1}\left\Vert X\right\Vert \left\Vert T_{t}^{0}[Y]-Y\right\Vert .
\end{equation*}%
Let $X,Y\in \mathrm{dom\,}\mathcal{L}_{0}.$ Then by (\ref{iii}) $XY\in
\mathrm{dom\,}\mathcal{L}_{0}.\square $

The perturbed backward Markovian master equation (\ref{bme}) is equivalent
to the integral equation%
\begin{equation}
\langle \varphi |T_{t}[X]|\psi \rangle =\langle \varphi |T_{t}^{0}[X]|\psi
\rangle +\int_{0}^{t}\,\mathrm{Tr}\,\Lambda \lbrack S_{s}^{0}\left[ |\psi
\rangle \langle \varphi |\right] ]T_{t-s}[X]\,ds,\quad \varphi ,\psi \in
\mathcal{D},  \label{ime}
\end{equation}%
see \cite{chf}, \cite{hol9} (Appendix). Taking into account (\ref{S01}), (%
\ref{lambda}), we have $\Lambda \lbrack S_{s}^{0}\left[ |\psi \rangle
\langle \varphi |\right] ]=\Omega \,\psi (s)\,\overline{\varphi (s)},$ so
that the integral term becomes $\int_{0}^{t}\,\,\psi (s)\,\overline{\varphi
(s)\,}\mathrm{Tr}\,\Omega T_{t-s}[X]ds.$ It follows%
\begin{equation}
\langle \varphi |t^{-1}\left( T_{t}[X]-X\right) |\psi \rangle =\langle
\varphi |t^{-1}\left( T_{t}^{0}[X]-X\right) |\psi \rangle  \label{ed}
\end{equation}%
\begin{equation*}
+t^{-1}\int_{0}^{t}\,\psi (s)\,\overline{\varphi (s)\,}\mathrm{Tr}\,\Omega
T_{t-s}[X]\,ds,\quad \varphi ,\psi \in \mathcal{D}.
\end{equation*}

\begin{lemma}
\label{l2} If $X\in \mathrm{dom\,}\mathcal{L}$ then $X_{0}=X-\left( \mathrm{%
Tr\,}\Omega X\right) I\in \mathrm{dom\,}\mathcal{L}_{0}$ and $\mathcal{L}[X]=%
\mathcal{L}_{0}[X_{0}].$
\end{lemma}

\emph{Proof.} Since $I\in \mathrm{dom\,}\mathcal{L}$ and $\mathcal{L}[I]=0,$
then $X_{0}\in \mathrm{dom\,}\mathcal{L}$ and (\ref{ed}) implies
\begin{equation*}
\langle \varphi |t^{-1}\left( T_{t}^{0}[X_{0}]-X_{0}\right) |\psi \rangle
=\langle \varphi |t^{-1}\left( T_{t}[X_{0}]-X_{0}\right) |\psi \rangle
\end{equation*}%
\begin{equation*}
-t^{-1}\int_{0}^{t}\,\psi (s)\,\overline{\varphi (s)\,}\mathrm{Tr}\,\Omega
\,t^{-1}\left( T_{t-s}[X_{0}]-X_{0}\right) \,ds,\quad \varphi ,\psi \in
\mathcal{D},
\end{equation*}%
From the criterion (\ref{iii}),
\begin{equation*}
\left\vert \langle \varphi |t^{-1}\left( T_{t}^{0}[X_{0}]-X_{0}\right) |\psi
\rangle \right\vert \end{equation*}\begin{equation*}\leq M\left\Vert \varphi \right\Vert \left\Vert \psi
\right\Vert +\left\vert \int_{0}^{t}\,\psi (s)\,\overline{\varphi (s)\,}%
M\,t^{-1}\left( t-s\right) ds\right\vert \leq 2M\left\Vert \varphi
\right\Vert \left\Vert \psi \right\Vert .
\end{equation*}
By the same criterion $X_{0}\in \mathrm{dom\,}\mathcal{L}_{0}$ .$\square $

\emph{Proof of Theorem \ref{t1}}. Let $X\in \mathrm{dom\,}\mathcal{L}, X^{\ast }X\in \mathrm{dom\,}\mathcal{L},\,\,XX^{\ast }\in \mathrm{dom\,}%
\mathcal{L}$ , then%
\begin{eqnarray*}
X &=&\left[ X-\left( \mathrm{Tr\,}\Omega X\right) I\right] +\left( \mathrm{%
Tr\,}\Omega X\right) I=X_{0}+\left( \mathrm{Tr\,}\Omega X\right) I, \\
X^{\ast }X &=&\left[ X^{\ast }X-\left( \mathrm{Tr\,}\Omega X^{\ast }X\right)
I\right] +\left( \mathrm{Tr\,}\Omega X^{\ast }X\right) I=Y_{0}+\left(
\mathrm{Tr\,}\Omega X^{\ast }X\right) I,
\end{eqnarray*}%
where $X_{0},Y_{0}\in \mathrm{dom\,}\mathcal{L}_{0}$ by Lemma \ref{l2}. Then%
\begin{equation*}
Y_{0}+\left( \mathrm{Tr\,}\Omega X^{\ast }X\right) I=\left[ X_{0}+\left(
\mathrm{Tr\,}\Omega X\right) I\right] ^{\ast }\left[ X_{0}+\left( \mathrm{%
Tr\,}\Omega X\right) I\right] ,
\end{equation*}%
whence%
\begin{equation*}
(\mathrm{Tr\,}\Omega X^{\ast }X-\left\vert \mathrm{Tr\,}\Omega X\right\vert
^{2})I=X_{0}^{\ast }X_{0}+X_{0}^{\ast }\left( \mathrm{Tr\,}\Omega X\right)
+\left( \overline{\mathrm{Tr\,}\Omega X}\right) X_{0}-Y_{0}.
\end{equation*}%
By Lemma \ref{l1}, the term in righthand side belongs to $\mathrm{dom\,}%
\mathcal{L}_{0}.$ Since $I\notin \mathrm{dom\,}\mathcal{L}_{0},$ this is
possible iff $\mathrm{Tr\,}\Omega X^{\ast }X=\left\vert \mathrm{Tr\,}\Omega
X\right\vert ^{2},$ i.e. the equality holds in the noncommutative
Cauchy-Schwarz inequality $\mathrm{Tr\,}\Omega X^{\ast }X\geq \left\vert
\mathrm{Tr\,}\Omega X\right\vert ^{2}.$ But this holds iff $\left[ X-\left(
\mathrm{Tr\,}\Omega X\right) I\right] \Omega =0.$ Applying similar argument
to $XX^{\ast },$ we obtain $$\Omega \left[ X-\left( \mathrm{Tr\,}\Omega
X\right) I\right] =0.\quad \square $$

\textit{Proof of Proposition \ref{p1} } Denote by $\mathcal{\tilde{K}}_{0}$
the operator acting as
\begin{equation*}
\mathcal{\tilde{K}}_{0}\omega (x,\,y)=\frac{d}{dt}\omega (x+t,y+t)|_{t=0},
\end{equation*}%
with the domain $\mathrm{dom\,}\mathcal{\tilde{K}}_{0}$ described in the
Proposition \ref{p1}. We will check that $\mathcal{\tilde{K}}_{0}$ satisfies
\begin{equation}
\left( \lambda \mathcal{I-\tilde{K}}_{0}\right) \mathcal{R}_{\lambda }\omega
=\omega ,\quad \omega \in \mathfrak{T}(\mathcal{H});\quad \mathcal{R}%
_{\lambda }\left( \lambda \mathcal{I-\tilde{K}}_{0}\right) \omega =\omega
,\quad \omega \in \mathrm{dom\,}\mathcal{\tilde{K}}_{0},  \label{rr}
\end{equation}
where $\mathcal{R}_{\lambda }$ is the resolvent of the semigroup $S_{t}^{0}$.
Then the first equality implies that $\mathcal{\tilde{K}}_{0}\supseteq
\mathcal{K}_{0},$ while the second -- that $\mathrm{dom\,}\mathcal{\tilde{K}}%
_{0}\subseteq \mathrm{dom\,}\mathcal{K}_{0},$ because then $\omega =\mathcal{%
R}_{\lambda }\left( \lambda \omega -\mathcal{\tilde{K}}_{0}\omega \right)
\in \mathrm{dom\,}\mathcal{K}_{0}.$ Thus $\mathcal{\tilde{K}}_{0}=\mathcal{K}%
_{0}.$

Let $\omega \in \mathfrak{T}(\mathcal{H}),$ then $\mathcal{R}_{\lambda
}[\omega ]\in \mathrm{dom\,}\mathcal{K}_{0},$ and
\begin{equation*}
\mathcal{R}_{\lambda }[\omega ](x,y)=\int_{0}^{\infty }\mathrm{e}^{-\lambda
t}\omega (x+t,y+t)dt.
\end{equation*}%
Notice that for any $\omega \in \mathfrak{T}(\mathcal{H})$ the function $%
\omega (x+t,y+t)$ is integrable in $t$ for almost all $(x,y)$. This follows
from the fact that this function is the diagonal value of the kernel of the
trace-class operator $W_{x}\omega W_{y}^{\ast }.$ For the same reason, $%
\mathcal{R}_{\lambda }[\omega ](x+t,y+t)\ $\ is integrable in $t$ for almost
all $(x,y)$.

To prove the first equality we compute the generalized derivative of $%
\mathcal{R}_{\lambda }[\omega ](x+t,y+t)$ with respect to $t.$ For any
smooth function $f(t)$ wth compact support
\begin{eqnarray*}
&&-\int_{0}^{\infty }\mathcal{R}_{\lambda }[\omega ](x+s,y+s)f^{\prime }(s)ds
\\
&=&-\int_{0}^{\infty }\int_{0}^{\infty }\mathrm{e}^{-\lambda t}\omega
(x+t+s,y+t+s)f^{\prime }(s)dsdt \\
&=&-\int_{0}^{\infty }\mathrm{e}^{\lambda s}\int_{s}^{\infty }\mathrm{e}%
^{-\lambda \xi }\omega (x+\xi ,y+\xi )d\xi \,f^{\prime }(s)ds \\
&=&\int_{0}^{\infty }\left[ -\omega (x+s,y+s)+\lambda \int_{0}^{\infty }%
\mathrm{e}^{-\lambda t}\omega (x+t+s,y+t+s)dt\right] \,f(s)ds,
\end{eqnarray*}%
where the last equality is obtained by integration by parts. It follows that
$\mathcal{R}_{\lambda }[\omega ](x+s,y+s)$ is absolutely continuous in $s$
and its derivative is equal to the expression in the squared brackets \cite%
{nik}, hence
\begin{equation*}
\mathcal{\tilde{K}}_{0}\mathcal{R}_{\lambda }[\omega ](x,y)=\frac{d}{dt}%
\mathcal{R}_{\lambda }[\omega ](x+t,y+t)|_{t=0}=\lambda \mathcal{R}_{\lambda
}[\omega ](x,y)-\omega (x,y),
\end{equation*}%
and the function on the left is kernel of a trace class operator, which
proves the first equality. The second equality follows similarly from
integration by parts:%
\begin{equation*}
\mathcal{R}_{\lambda }\mathcal{\tilde{K}}_{0}\omega (x,y)=\int_{0}^{\infty }%
\mathrm{e}^{-\lambda t}\frac{d}{dt}\omega (x+t,y+t)dt=\lambda \mathcal{R}%
_{\lambda }[\omega ](x,y)-\omega (x,y).
\end{equation*}%
$\square $

\section{A nonstandard dynamical semigroup}

\label{s2}

\subsection{Quantum diffusion with extinction}

Let $\mathcal{H=}$ $L^{2}(\mathbb{R}_{+}),$ and trace-class operators $%
\omega $ in $\mathcal{H}$ are identified with their kernels $\omega
(x,y),\,x,y\in \mathbb{R}_{+}.$ Let $\mathcal{D}=AC_{0}^{2}(\mathbb{R}_{+})$
be the subspace of differentiable functions $%
\varphi (x)$ on $\mathbb{R}_{+}$ with $\varphi (0)=0,$ and such that $%
\varphi ^{\prime }(x)$ is absolutely continuous with $\varphi ^{\prime
\prime }\in $ $L^{2}(\mathbb{R}_{+}).$ Notice that $\varphi ^{\prime }(0)$
exists and is finite for $\varphi \in AC_{0}^{2}(\mathbb{R}_{+})$.

Consider the operators $L=\sqrt{2}\frac{d}{dx}$ with $\mathcal{D}\left(
L\right) =\mathcal{D},\,L^{\ast }=-\sqrt{2}\frac{d}{dx},\,$and $K=-\frac{%
d^{2}}{dx^{2}}$ selfadjoint with the domain $\mathcal{D}.$ The condition (%
\ref{1}) is then fulfilled with equality. The corresponding backward MME (%
\ref{2}), in which the sum consists of one term, has the form
\begin{equation}
\frac{d}{dt}\langle \varphi |T_{t}[X]|\psi \rangle =2\langle \varphi
^{\prime }|T_{t}[X]|\psi ^{\prime }\rangle +\langle \varphi ^{\prime \prime
}|T_{t}[X]|\psi \rangle +\langle \varphi |T_{t}[X]|\psi ^{\prime \prime
}\rangle ,\quad \varphi ,\psi \in \mathcal{D}.  \label{Sf}
\end{equation}%
The conditions of Theorems A.1, A.2 in \cite{hol9} are fulfilled in this
case ensuring existence of the minimal solution of the MME which is a
standard nonunital dynamical semigroup $T_{t}^{0},$ with the predual
semigroup $S_{t}^{0}$ in $\mathfrak{T}(\mathcal{H})$ satisfying the forward
MME, which in this special case has the form similar to (\ref{Sf}):%
\begin{equation}
\frac{d}{dt}\langle f|S_{t}^{0}[\omega ]|g\rangle =2\langle f^{\prime
}|S_{t}^{0}[\omega ]|g^{\prime }\rangle +\langle f^{\prime \prime
}|S_{t}^{0}[\omega ]|g\rangle +\langle f|S_{t}^{0}[\omega ]|g^{\prime \prime
}\rangle ,\quad f,g\in \mathcal{D}.  \label{3a}
\end{equation}%
The generator $\mathcal{K}_{0}$ of $S_{t}^{0}$ is an extension of the
operator
\begin{equation}
\mathcal{K}_{0}\left[ \omega \right] =\omega _{xx}^{\prime \prime
}(x,y)+2\omega _{xy}^{\prime \prime }(x,y)+\omega _{yy}^{\prime \prime
}(x,y),  \label{K0}
\end{equation}%
defined initially on the domain
\begin{equation*}
\mathfrak{D}_{0}=\mathrm{lin}\left\{ \omega :\omega =|\psi \rangle \langle
\varphi |;\,\varphi ,\psi \in \mathcal{D}\right\} \,\mathfrak{\subset T}(%
\mathcal{H}),
\end{equation*}%
see Example 3 in \cite{hol4}. A detailed description of $\mathrm{\,}\mathcal{%
K}_{0}$ and its domain is given in Proposition \ref{p3} in the subsection %
\ref{s33}.

\begin{lemma}
\label{l3} Let $\varphi ,\psi \in \mathcal{H}$ be such that $|\psi \rangle
\langle \varphi |\,\in \mathrm{dom\,}\mathcal{K}_{0},$ then $|\psi \rangle
,\,|\varphi \rangle \subseteq AC_{0}^{2}(\mathbb{R}_{+}),$ in particular $%
\psi (0)=\varphi (0)=0,$ and
\begin{equation}
\mathcal{K}_{0}[|\psi \rangle \langle \varphi |](x,y)=2\psi ^{\prime }(x)%
\overline{\varphi ^{\prime }(y)}+\psi ^{\prime \prime }(x)\overline{\varphi
(y)}+\psi (x)\overline{\varphi ^{\prime \prime }(y)}.  \label{phipsi}
\end{equation}
\end{lemma}

\textit{Proof.} If $\omega \in \mathrm{dom\,}\mathcal{K}_{0}$ then taking $%
t=0$ in (\ref{3a}) we obtain
\begin{equation*}
\langle f|\mathcal{K}_{0}[\omega ]|g\rangle =2\langle f^{\prime }|\omega
|g^{\prime }\rangle +\langle f^{\prime \prime }|\omega |g\rangle +\langle
f|\omega |g^{\prime \prime }\rangle ,\quad f,g\in AC_{0}^{2}(\mathbb{R}_{+}).
\end{equation*}%
For $\omega =|\psi \rangle \langle \varphi |$ this amounts to
\begin{equation*}
\langle f|\mathcal{K}_{0}[\omega ]|g\rangle =2\langle f^{\prime }|\psi
\rangle \langle \varphi |g^{\prime }\rangle +\langle f^{\prime \prime }|\psi
\rangle \langle \varphi |g\rangle +\langle f|\psi \rangle \langle \varphi
|g^{\prime \prime }\rangle .
\end{equation*}%
Take $g$ such that $\langle \varphi |g^{\prime }\rangle =0,\langle \varphi
|g\rangle =1$ (this is possible because $g$ and $g^{\prime }$ are linearly
independent for $g\in AC_{0}^{2}(\mathbb{R}_{+})$), then%
\begin{equation*}
\langle f^{\prime \prime }|\psi \rangle =\langle f|g_{1}\rangle ,\quad f\in
AC_{0}^{2}(\mathbb{R}_{+}),
\end{equation*}%
where $|g_{1}\rangle =\mathcal{K}_{0}[\omega ]|g\rangle -|\psi \rangle
\langle \varphi |g^{\prime \prime }\rangle \in L^{2}(\mathbb{R}_{+}).$
Restricting to infinite differentiable functions $f$ with compact support in
$(0,\infty ),$ this means that $\psi $ has generalized second derivative in $%
(0,\infty )$, see \cite{nik}, which belongs to $L^{2}(\mathbb{R}_{+}).$
Moreover, for continuously differentiable $f$ \ in $\mathbb{R}_{+}$ one can
integrate by parts, obtaining%
\begin{equation*}
f\,^{\prime }(0)\psi (0)-\int_{0}^{\infty }f\,^{\prime }(x)\psi ^{\prime
}(x)dx=\int_{0}^{\infty }f\,(x)g_{1}(x)dx.
\end{equation*}%
One can choose $f$ such that $f\,^{\prime }(0)\neq 0,$ while both integrals
are arbitrarily small. Therefore $\psi (0)=0,$ hence $\psi \in AC_{0}^{2}(%
\mathbb{R}_{+}).$ Similar proof applies to $\varphi $. Relation (\ref{phipsi}%
) follows from (\ref{K0}). $\square $

\subsection{Quantum diffusion with rebound}

The semigroup $S_{t}^{0}$ describes \textquotedblleft noncommutative
diffusion on $\mathbb{R}_{+}$ with absorption at the point
0\textquotedblright\ (with extinction of the absorbed particle). The example
of the nonstandard semigroup will be obtained as a result of perturbation of
the generator of this semigroup by the term $\Lambda \left[ \omega \right]
=\Omega \frac{d}{dx}\omega (x,x)|_{x=0},$ where $\Omega $ is a fixed density
operator. Consider the perturbed generator
\begin{equation}
\mathcal{K}\left[ \omega \right] =\mathcal{K}_{0}\left[ \omega \right]
+\Omega \frac{d}{dx}\omega (x,x)|_{x=0}.\quad  \label{K}
\end{equation}%
which corresponds to \textquotedblleft rebound from 0 to the state $\Omega $
\textquotedblright\ \cite{hol4}. Since the perturbation has rank one, it
follows (see \cite{hol4})
\begin{equation}
\mathrm{dom\,}\mathcal{K}=\mathrm{dom\,}\mathcal{K}_{0}.  \label{dd}
\end{equation}%
The semigroup $T_{t}$ is the minimal solution of the MME%
\begin{equation*}
\frac{d}{dt}\mathrm{Tr\,}\omega T_{t}[X]=\mathrm{Tr\,}\mathcal{K\,}[\omega
\,]T_{t}[X],\quad \omega \in \mathrm{dom\,}\mathcal{K}_{0},
\end{equation*}%
and it is unital \cite{hol4}.

We will prove that $T_{t}$\textit{\ is not standard.}

Assume the contrary, i.e. that there exist some operators $K,L_{j}$ defined
on $\mathrm{dom\,}K$ (such that $K$ is $m$-accretive) and satisfying (\ref{1}%
) such that $T_{t}$ is the minimal solution of the Eq. (\ref{2}), or
equivalently (\ref{2a}).

\begin{lemma}
\label{l4}
\begin{equation}
\mathcal{K\,}[|\psi \rangle \langle \varphi |\,]=\mathcal{K}_{0}\mathcal{\,}%
[|\psi \rangle \langle \varphi |\,],\quad \varphi ,\,\psi \in \mathrm{dom\,}%
K.  \label{5a}
\end{equation}
\end{lemma}

\textit{Proof. } If $\varphi ,\,\psi \in \mathrm{dom\,}K,$ then $|\psi
\rangle \langle \varphi |\,\in \,\mathrm{dom\,}\mathcal{K}$ (see Example 1
in \S\ 4 of \cite{hol4}), hence by (\ref{dd}) $|\psi \rangle \langle \varphi
|\,\in \mathcal{\,}\mathrm{dom\,}\mathcal{K}_{0}.$ By Lemma \ref{l3} we then
have $\psi ,\varphi \in AC_{0}^{2}(\mathbb{R}_{+}).$ Thus
\begin{equation}
\mathrm{dom\,}K\subseteq AC_{0}^{2}(\mathbb{R}_{+}).  \label{inc}
\end{equation}%
The perturbation term\textit{\ }in (\ref{K}) vanishes for $\omega
(x,x)=\psi (x)\overline{\varphi (x)}$ because
\begin{equation*}
\frac{d}{dx}\psi (x)\overline{\varphi (x)}|_{x=0}=\psi ^{\prime }(0)%
\overline{\varphi (0)}+\psi (0)\overline{\varphi ^{\prime }(0)}=0
\end{equation*}%
and $\psi (0)=\varphi (0)=0$ due to (\ref{inc}) $.\square $

\begin{lemma}
\label{l5} $\mathrm{dom\,}K=AC_{0}^{2}(\mathbb{R}_{+}).$
\end{lemma}

\textit{Proof}. The equation (\ref{2}) is invariant under transformations%
\begin{equation*}
L_{j}\rightarrow L_{j}^{\prime }=L_{j}+\alpha _{j},\quad K\rightarrow
K^{\prime }=K+\sum_{j}\left( \bar{\alpha}_{j}L_{j}+\frac{\left\vert \alpha
_{j}\right\vert ^{2}}{2}\right) ,
\end{equation*}%
where $\sum_{j}\left\vert \alpha _{j}\right\vert ^{2}<\infty .$ Then the $m$%
-accretive operator $K$ is transformed into the $m$-accretive operator $%
K^{\prime }$ with the same domain
\begin{equation}
\mathrm{dom\,}K^{\prime }=\mathrm{dom\,}K,  \label{dodo}
\end{equation}%
as it follows from the Corollary to Theorem X.50 of \cite{rs2} \ and the
following estimate
\begin{equation}
\left\Vert \sum_{j}\bar{\alpha}_{j}L_{j}\psi \right\Vert \leq \frac{1}{\sqrt{%
2}}\left\Vert \left( K+\sum_{j}\frac{\left\vert \alpha _{j}\right\vert ^{2}}{%
2}\right) \psi \right\Vert +\frac{\sum_{j}\left\vert \alpha _{j}\right\vert
^{2}}{2\sqrt{2}}\left\Vert \psi \right\Vert ,\quad \psi \in \mathrm{dom\,}K.
\label{6}
\end{equation}%
To prove this estimate, observe that%
\begin{eqnarray}
\left\Vert \sum_{j}\bar{\alpha}_{j}L_{j}\psi \right\Vert ^{2} &\leq
&\sum_{j}\left\vert \alpha _{j}\right\vert ^{2}\sum_{j}\left\Vert L_{j}\psi
\right\Vert ^{2}  \label{7} \\
&\leq &2\mathrm{Re}\,\sum_{j}\left\vert \alpha _{j}\right\vert
^{2}\left\langle \psi \,|K\,\psi \right\rangle \leq \frac{1}{2}\left\Vert
\left( K+\sum_{j}\left\vert \alpha _{j}\right\vert ^{2}\right) \psi
\right\Vert ^{2},  \notag
\end{eqnarray}%
where in the second inequality we used (\ref{1}), and in the last -- the
general relation%
\begin{equation*}
2\mathrm{Re}\,\left\langle \psi \,|\,\varphi \right\rangle =\frac{1}{2}%
\left\Vert \psi +\varphi \right\Vert ^{2}-\frac{1}{2}\left\Vert \psi
-\varphi \right\Vert ^{2}\leq \frac{1}{2}\left\Vert \psi +\varphi
\right\Vert ^{2}.
\end{equation*}%
Then (\ref{6}) follows from (\ref{7}) by splitting $K+\sum_{j}\left\vert
\alpha _{j}\right\vert ^{2}=\left( K+\sum_{j}\left\vert \alpha
_{j}\right\vert ^{2}/2\right) +\sum_{j}\left\vert \alpha _{j}\right\vert
^{2}/2$ and using the triangle inequality.

By using (\ref{5a}) we obtain
\begin{equation*}
\sum_{j}|L_{j}^{\prime }\psi \rangle \langle L_{j}^{\prime }\varphi
|\,-\,|K^{\prime }\,\psi \rangle \langle \varphi |\,-|\,\psi \rangle \langle
K^{\prime }\varphi |\,=\mathcal{K\,}[|\psi \rangle \langle \varphi |\,]=%
\mathcal{K}_{0}[|\psi \rangle \langle \varphi |\,],\quad \varphi ,\,\psi \in
\mathrm{dom\,}K.
\end{equation*}%
Let us fix a unit vector $\psi _{0}\in \mathrm{dom\,}K$ and put $\alpha
_{j}=-\left\langle \psi _{0}\,|L_{j}|\psi _{0}\right\rangle ,$ then $%
\left\langle \psi _{0}\,|L_{j}^{\prime }|\psi _{0}\right\rangle =0$. Taking $%
\varphi =\psi _{0}$ and multiplying by $|\psi _{0}\rangle ,$ we obtain%
\begin{equation}
K^{\prime }\,\psi (x)=-\psi (x)\langle K^{\prime }\psi _{0}|\psi _{0}\rangle
-\int_{0}^{\infty }\mathcal{K}_{0}\mathcal{\,}[|\psi \rangle \langle \psi
_{0}|](x,y)\,\,\psi _{0}(y)\,dy  \label{int}
\end{equation}%
for $\psi \in \mathrm{dom\,}K.$ Notice that $\psi _{0}(0)=0$ since $\psi
_{0}\in \mathrm{dom\,}K.$ By using (\ref{phipsi}), we have
\begin{equation*}
\mathcal{K}_{0}\mathcal{\,}[|\psi \rangle \langle \psi _{0}|(x,y)=\,\psi
^{\prime \prime }(x)\,\overline{\psi _{0}(y)}+2\psi ^{\prime }(x)\,\overline{%
\psi _{0}^{\prime }(y)}\,+\psi (x)\,\overline{\psi _{0}^{\prime \prime }(y)}
\end{equation*}%
hence%
\begin{equation*}
\int_{0}^{\infty }\mathcal{K}_{0}\mathcal{\,}[|\psi \rangle \langle \psi
_{0}|(x,y)\,\psi _{0}(y)\,dy=\,\psi ^{\prime \prime }(x)+2i\,c_{1}\psi
^{\prime }(x)+c_{0}\psi (x),
\end{equation*}%
where%
\begin{equation*}
c_{1}=-i\int_{0}^{\infty }\overline{\psi _{0}^{\prime }(y)}\,\psi
_{0}(y)dy\in \mathbb{R},\end{equation*}\begin{equation*}c_{0}=\int_{0}^{\infty }\overline{\psi
_{0}^{\prime \prime }(y)}\,\psi _{0}(y)dy=-\int_{0}^{\infty }\left\vert \psi
_{0}^{\prime }(y)\right\vert ^{2}dy\leq 0.
\end{equation*}
Here in integration by parts we took into account that $\psi _{0}(0)=0.$
Therefore the righthand side of (\ref{int}), up to an additive term which is
a multiple of unit operator, defines an accretive operator of the form $%
-\,\psi ^{\prime \prime }-2ic_{1}\psi ^{\prime }+c_{2}\psi $ for $\psi \in
AC_{0}^{2}(\mathbb{R}_{+}).$ By the gauge transformation $\psi
(x)\rightarrow \psi (x)e^{-2ic_{1}x},$ leaving $AC_{0}^{2}(\mathbb{R}_{+})$
invariant, this operator is unitarily equivalent to $-\,\psi ^{\prime \prime
}+c_{3}\psi ,$ therefore it is $m$-accretive. Since $K$ is $m$-accretive, $%
\mathrm{dom\,}K^{\prime }=$ $\mathrm{dom\,}K=AC_{0}^{2}(\mathbb{R}_{+})$ and
$K^{\prime }=-\,\psi ^{\prime \prime }+c_{3}\psi .$\,$\square $

We have assumed that $T_{t}$ is standard. To obtain a contradiction, notice
that Lemmas \ref{l4} and \ref{l5} imply that $\mathcal{K\,}[|\psi \rangle
\langle \varphi |\,]=\mathcal{K}_{0}\mathcal{\,}[|\psi \rangle \langle
\varphi |\,]$ for $\varphi ,\,\psi \in \mathcal{D}=\mathrm{dom\,}%
K=AC_{0}^{2}(\mathbb{R}_{+}).$ Thus both $T_{t}$ and $T_{t}^{0}$ are the
minimal solutions of the same Eq. (\ref{2a}), hence $\mathcal{K}\,=\,\mathcal{K%
}_{0}$ which is a contradiction.$\,\square $

\textbf{Remark}
 We conclude with a brief comment concerning \textquotedblleft
nonstandardness\textquotedblright\ of the perturbation term
\begin{equation*}
\Lambda \mathcal{\,}[\omega \,]=\Omega \frac{d}{dx}\omega (x,x)|_{x=0}.
\end{equation*}
Its complete positivity as a map from $\mathrm{dom\,}\mathcal{K}_{0}$ to $%
\mathfrak{T}(\mathcal{H})$ amounts to the positivity of the unbounded linear
functional $\ \omega \rightarrow \frac{d}{dx}\omega (x,x)|_{x=0},$ defined
on $\mathrm{dom\,}\mathcal{K}_{0}.$ Indeed, $\omega \geq 0$ implies $\omega
(x,x)\geq 0,$ and if $\omega \in \mathrm{dom\,}\mathcal{K}_{0},$ then $%
\omega (x,0)=\omega (0,y)=0,$ hence $\omega (0,0)=0$ and necessarily $\frac{d%
}{dx}\omega (x,x)|_{x=0}\geq 0.$ But the functional $\omega \rightarrow
\frac{d}{dx}\omega (x,x)|_{x=0}$ has meaning for a broader domain of
trace-class operators $\omega ,$ for which the kernel $\omega (x,y)$ need
not satisfy the zero boundary condition, and then $\omega \geq 0$ need not
imply $\frac{d}{dx}\omega (x,x)|_{x=0}\geq 0.$ An example is $\omega
(x,y)=\psi (x)\overline{\psi (y)}$, where $\psi $ is a square-integrable
with $\mathrm{Re}\,\psi ^{\prime }(0)\overline{\psi (0)}<0$ (e.g. $\psi
(x)=\exp (-x).$) Formally
\begin{equation*}
\frac{d}{dx}\omega (x,x)|_{x=0}=-\langle \delta ^{\prime }|\omega |\delta
\rangle -\langle \delta |\omega |\delta ^{\prime }\rangle ;
\end{equation*}%
for this reason the perturbation term $\Lambda \mathcal{\,}[\omega \,]$ does
not have any generalization of the Kraus form for CP maps.

\subsection{The semigroup $S_{t}^{0}$}

\label{s33}

Here we give an explicit expression for $S_{t}^{0}$ which however is not
needed for the proof of nonstandardness. From (\ref{3a}) it follows that the
family $\omega _{t}=S_{t}^{0}[\omega ]$ satisfies
\begin{equation}
\frac{d}{dt}\langle f|\omega _{t}|g\rangle =2\langle f^{\prime }|\omega
_{t}|g^{\prime }\rangle +\langle f^{\prime \prime }|\omega _{t}|g\rangle
+\langle f|\omega _{t}|g^{\prime \prime }\rangle ,\quad f,g\in \mathcal{D}.
\label{Sff}
\end{equation}

Consider the expression (\ref{K0}) for the generator $\mathcal{K}_{0}$ of
the semigroup $S_{t}^{0}.$ Making change of variables $u=x+y,\,v=x-y,$ one
can see that $\mathcal{K}_{0}$ is a restriction to $\mathfrak{D}_{0}$ of the
operator
\begin{equation}
\mathcal{K}_{0}\left[ \omega \right] =4\frac{\partial ^{2}}{\partial u^{2}}%
\omega \left( \frac{u+v}{2},\frac{u-v}{2}\right) =\frac{d^{2}}{d\xi ^{2}}%
\omega (x+\xi ,y+\xi )|_{\xi =0}  \label{K0hat}
\end{equation}%
of the second directional derivative at the point $(x,y)$ in the direction $%
(1,1),$ defined on a broader domain \textrm{dom }$\mathcal{K}_{0}.$

The quadrant $Q=\mathbb{R}_{+}\times \mathbb{R}_{+}$ in the variables $u,\,v$
is given by the inequality $u\geq |v|.$ Consider the Cauchy problem for the
degenerate heat equation%
\begin{equation}
\frac{\partial \omega _{t}}{\partial t}=4\frac{\partial ^{2}\omega _{t}}{%
\partial u^{2}},\quad u>|v|,\quad \omega _{t}|_{\partial Q}=0,  \label{cau}
\end{equation}%
in which the variable $v$ enters as a parameter. The solution is obtained by
the method of reflection. For given $x,y,$ define $\tilde{\omega}_{0}(\xi +%
\left[ x-y\right] _{+},\xi +\left[ y-x\right] _{+})=-\omega _{0}(-\xi +\left[
x-y\right] _{+},-\xi +\left[ y-x\right] _{+})$ \ for $\xi \leq 0$ -- the odd
continuation (with respect to the point $\xi =0$) of $\omega _{0}$ along the
line $u_{0}=|x-y|+2\xi ,\,v_{0}=x-y;\,\xi \in \mathbb{R}.$ Then the solution
of (\ref{cau}) is given by the Poisson integral
\begin{eqnarray}
\omega _{t}(x,y) &\equiv &\omega _{t}\left( \frac{u+v}{2},\frac{u-v}{2}%
\right)  \notag \\
&=&\frac{1}{4\sqrt{\pi t}}\int_{-\infty }^{\infty }\exp \left\{ -\frac{%
|u-u_{0}|^{2}}{16t}\right\} \tilde{\omega}_{0}\left( \frac{u_{0}+v}{2},\frac{%
u_{0}-v}{2}\right) du_{0}  \notag \\
&=&\frac{1}{2\sqrt{\pi t}}\int_{0}^{\infty }\sum\limits_{n=0,1}\left(
-1\right) ^{n}\exp \left\{ -\frac{|\min (x,y)-\left( -1\right) ^{n}\xi |^{2}%
}{4t}\right\}  \notag \\
&\times &\omega _{0}(\xi +\left[ x-y\right] _{+},\xi +\left[ y-x\right]
_{+})d\xi .  \label{S0}
\end{eqnarray}%
The expression (\ref{S0}) should replace the incorrect unnumbered formula at the
bottom of p.1 in \cite{hol5}.

\begin{proposition}
\label{p2} The Cauchy problem (\ref{cau}) is equivalent to the Cauchy
problem for the equation (\ref{Sff}).
\end{proposition}

\textit{Proof (Sketch). }Denoting $F(x,y)=$\textit{\ }$g(y)$ $\overline{f(x)}
$ we have
\begin{eqnarray*}
\langle f|\omega _{t}|g\rangle &=&\int_{0}^{\infty }\int_{0}^{\infty }\omega
_{t}(x,y)F(x,y)dxdy \\
&=&\int \int_{Q}\omega _{t}\left( \frac{u+v}{2},\frac{u-v}{2}\right) F\left(
\frac{u+v}{2},\frac{u-v}{2}\right) dudv.
\end{eqnarray*}%
From (\ref{cau}) we obtain%
\begin{equation}
\frac{d}{dt}\langle f|\omega _{t}|g\rangle =4\int \int_{Q}\left[ \frac{%
\partial ^{2}}{\partial u^{2}}\omega _{t}\left( \frac{u+v}{2},\frac{u-v}{2}%
\right) \right] F\left( \frac{u+v}{2},\frac{u-v}{2}\right) dudv.
\label{Sfff}
\end{equation}%
Integrating by parts, the righthand side is equal to%
\begin{equation*}
4\int \int_{Q}\omega _{t}\left( \frac{u+v}{2},\frac{u-v}{2}\right) \frac{%
\partial ^{2}}{\partial u^{2}}F\left( \frac{u+v}{2},\frac{u-v}{2}\right)
dudv+4\int_{\partial Q}\omega _{t}\frac{\partial }{\partial u}F\,ds.
\end{equation*}%
Taking into account zero boundary values of $\omega _{t}$ on $\partial Q,$
we obtain%
\begin{eqnarray}
\frac{d}{dt}\langle f|\omega _{t}|g\rangle &=&4\int \int_{Q}\omega
_{t}\left( \frac{u+v}{2},\frac{u-v}{2}\right) \frac{\partial ^{2}}{\partial
u^{2}}F\left( \frac{u+v}{2},\frac{u-v}{2}\right) dudv  \label{W1} \\
&=&2\int_{0}^{\infty }\int_{0}^{\infty }\overline{f^{\prime }(x)}\omega
_{t}(x,y)g^{\prime }(y)dxdy  \notag \\
&&+\int_{0}^{\infty }\int_{0}^{\infty }\overline{f^{\prime \prime }(x)}%
\omega _{t}(x,y)g(y)dxdy+\int_{0}^{\infty }\int_{0}^{\infty }\overline{f(x)}%
\omega _{t}(x,y)g^{\prime \prime }(y)dxdy  \notag
\end{eqnarray}%
whence (\ref{Sff}) follows.

Conversely, starting from (\ref{Sff}) we obtain (\ref{W1}). Since the linear
span of the products $F(x,y)=g(y)\overline{f(x)}$ is dense in the space of \
test functions, this means that%
\begin{equation*}
\frac{d}{dt}\int \int_{Q}\omega _{t}\,F\,dudv=4\int \int_{Q}\omega _{t}\frac{%
\partial ^{2}}{\partial u^{2}}F\,dudv
\end{equation*}%
for all test functions $F$ in $Q.$ Taking $F$ as a product $a(u)\,b(v),$ we
obtain that for almost all $v$ the function $\omega _{t}$ is the weak
solution of the heat equation (\ref{cau}). But the weak solution of the heat
equation coincides with the classical solution \cite{vlad}, hence we can
perform partial integration to obtain
\begin{equation}
\int_{\partial Q}\omega _{t}\frac{\partial }{\partial u}F\,ds=0  \label{gr}
\end{equation}%
for all $F\in \mathfrak{D}_{0}.$ Taking $F(x,y)=g(y)\overline{f(x)},$ where $%
f,g\in AC_{0}^{2}(\mathbb{R}_{+}),$ we obtain
\begin{equation*}
\left. \frac{\partial }{\partial u}\right\vert _{\partial Q}F\,=\frac{1}{2}%
\left\{
\begin{array}{c}
g^{\prime }(0)\overline{f(x)},\quad x>0,y=0, \\
g(y)\overline{f^{\prime }(0)}\quad x=0,y>0%
\end{array}%
\right. .
\end{equation*}
Therefore we can choose $f,g$ such that $\left. \frac{\partial }{\partial u}%
\right\vert _{\partial Q}F\,$\ is an arbitrary function from $AC_{0}^{2}(%
\mathbb{R}_{+}).$ From (\ref{gr}) it follows that $\left. \omega
_{t}\right\vert _{\partial Q}=0.$ Then we can modify $\omega _{t}$ by
changing it on a set of zero Lebesgue measure in $Q$ so that it will be a
solution of the Cauchy problem (\ref{cau}). $\square $

In this way we can also prove one part of the following statement:

\begin{proposition}
\label{p3} $\mathrm{dom\,}\mathcal{K}_{0}$ consists of trace-class operators
$\omega $ with kernels $\omega (x,y)$ such that for almost all $(x,y)$ the
function $\omega (x+\xi,y+\xi)$ as a function of $\xi$ is absolutely continuous,
vanishes for $\xi=0$ and has second generalized derivative $\sigma (x,y)=\frac{d^{2}}{d\xi ^{2}}\omega
(x+\xi ,y+\xi )|_{\xi =0}$, which is a kernel of a trace-class
operator in $\mathcal{H}$.
\end{proposition}

\textit{Proof (Sketch). } Indeed, if $\omega \in \mathrm{dom\,}\mathcal{K}%
_{0}$ then taking $t=0$ in (\ref{3a}) we obtain
\begin{equation*}
\langle f|\mathcal{K}_{0}[\omega ]|g\rangle =2\langle f^{\prime }|\omega
|g^{\prime }\rangle +\langle f^{\prime \prime }|\omega |g\rangle +\langle
f|\omega |g^{\prime \prime }\rangle ,\quad f,g\in AC_{0}^{2}(\mathbb{R}_{+}),
\end{equation*}%
which is equal to (\ref{W1}) for $t=0$. It follows that
\begin{equation*}
\int \int_{Q}\sigma Fdudv=4\int \int_{Q}\omega \frac{\partial ^{2}}{\partial
u^{2}}Fdudv,
\end{equation*}%
for arbitrary test function $F$, where $\sigma $ denotes kernel of the
trace-class operator $\mathcal{K}_{0}[\omega ]$. This implies that $\sigma $
is the generalized second derivative of $\omega $ (see \cite{nik}),
\begin{equation}
\sigma (x,y)=\mathcal{K}_{0}[\omega ](x,y)=\frac{d^{2}}{d\xi ^{2}}\omega
(x+\xi ,y+\xi )|_{\xi =0},\quad \omega \in \mathrm{dom\,}\mathcal{K}_{0}.
\label{K0K}
\end{equation}%
Moreover, integrating by parts and taking into account that $\frac{\partial
}{\partial u}F$ can be arbitrary, we obtain the boundary condition $\omega
|_{\partial Q}=0$. Thus $\mathcal{K}_{0}\subseteq \tilde{\mathcal{K}}_{0}$,
the operator (\ref{K0hat}) with the domain described in the Proposition \ref%
{p3}.

To prove converse inclusion, one can verify the resolvent relations (\ref{rr}%
) (in fact, only the second one) for the operator $\tilde{\mathcal{K}}_{0}$,
with the resolvent
\begin{equation*}
\mathcal{R}_{\lambda }[\omega ]=\int_{0}^{\infty }\mathrm{e}^{-\lambda
t}S_{t}^{0}[\omega ]dt
\end{equation*}%
computed according to (\ref{S0}), see \cite{hsw}. This can be done similarly to the proof of
Proposition \ref{p1}. $\square $

\textit{Acknowledgement.} The author acknowledges discussions with R.F. Werner which
stimulated revival of the author's interest to unbounded generators of quantum dynamical semigroup
and led to substantial clarifications of the mechanism of nonstandardness.
This work is supported by the Russian Science Foundation under grant 14-21-00162.


\begin{thebibliography}{99}
\bibitem{arv} W. Arveson, The domain algebra of a CP-semigroup, Pacific. J.
Math. 203, 1 67-77 (2002). 

\bibitem{alaz} S. Alazzawi, B. Baumgartner, Generalized Kraus operators and
generators of quantum dynamical semigroups, Arxiv:1306.4531

\bibitem{bhat} B. V. Bhat, K. R. Parthasarathy, Markov dilations of
nonconservative dynamical semigroups and quantum boundary theory, Ann. Inst.
H. Poincare, ser. B \textbf{31}, 601-652 (1995).

\bibitem{chebot} A. M. Chebotarev, Sufficient conditions for conservativity
of the minimal dynamical semigroups, Theor. Math. Phys. \textbf{80}, 192-211
(1989). 

\bibitem{chebot1} A. M. Chebotarev, Lectures on quantum probability,
Sociedad Matematica Mexicana, Textos 14, Mexico 2000.

\bibitem{chf} A. M. Chebotarev, F. Fagnola, Sufficient conditions for
conservativity of quantum dynamical semigroups, J. Funct. Anal. \textbf{118}%
, 131-153 (1993).

\bibitem{dav2} E. B. Davies, Quantum dynamical semigroups and the neutron
diffusion equation, Rep. Math. Phys. \textbf{11}, 169--188 (1977).

\bibitem{fel} W. Feller, An introduction to probability theory and its
applications, vol. I, II, John Wiley, NY.






\bibitem{gks} V. Gorini, A. Kossakowski, E. C. G. Sudarshan, Completely
positive dynamical semigroups of N-level systems, J. Math. Phys. \textbf{17}%
, 821-825 (1976).

\bibitem{hol4} A. S. Holevo, Excessive maps, ``arrival times'' and
perturbations of dynamical semigroups, Izvestiya: Mathematics \textbf{59}:6,
1311-1325 (1995).

\bibitem{hol3} A. S. Holevo, On the structure of covariant dynamical
semigroups, J. Funct. Anal. \textbf{131}, 255-278 (1995).


\bibitem{hol9} A. S. Holevo, On dissipative stochastic equations in a
Hilbert space, Probab. Theory Rel. Fields \textbf{104}, 483-500 (1996).

\bibitem{hol5} A. S. Holevo, There exists a nonstandard dynamical semigroup
on $\mathfrak{B}(\mathcal{H})$, Uspekhi Mat. Nauk. \textbf{51}(6), 225-226
(1996).



\bibitem{lindblad} G. Lindblad, On generators of quantum dynamical
semigroups, Commun. Math. Phys. \textbf{48}, 119--130 (1976).

\bibitem{sinha} A. Mohari, K. B. Sinha, Stochastic dilation of minimal
quantum dynamical semigroup, Proc. Indian Acad. Sci., \textbf{102}, 159-173
(1992).



\bibitem{nik} S.M. Nikol'sky, Approximation of functions of several
variables and embedding theorems, Nauka, Moscow 1977, \S 4.1.

\bibitem{rs2} M. Reed, B. Simon, Methods of modern mathematical physics. II.
Fourier analysis and self-adjointness, AP, NY 1975.

\bibitem{vlad} V. S. Vladimirov, Equations of mathematical physics, 5-th
edition, Nauka, Moscow,  1988.

\bibitem{hsw} I. Siemon, A. S. Holevo, R. F. Werner, Unbounded generators of
dynamical semigroups, Open Syst. Inf. Dyn. \textbf{24} (4), 1740015 (2017).









\end{thebibliography}
\end{document}